\documentclass[twocolumn]{aastex701}
\usepackage{amsmath}
\usepackage{longtable}
\usepackage{changepage}
\usepackage{rotating}
\usepackage{silence}
\usepackage{graphicx}
\usepackage{subcaption}
\usepackage{float}
\usepackage{xcolor}
\usepackage{hyperref}
\usepackage{booktabs}
\providecommand{\sw}[1]{\texttt{#1}}

\newcommand{\eg}{$\mathrm{erg~cm^{-2}}$}
\newcommand{\jetsimpy}{\sw{jetsimpy}}

\newcommand{\swift}{\emph{Swift}}
\newcommand{\fermi}{\emph{Fermi}}
\newcommand{\astrosat}{\emph{AstroSat}}

\newcommand{\thisgrb}{GRB\,250916A}
\newcommand{\tnt}{T\(_{90}\)}

\newcommand{\eiso}{\(E_{iso}\)}
\newcommand{\ekiso}{\(E_{k,{\rm iso}}\)}
\newcommand{\egmiso}{\(E_{\gamma,\textrm{iso}}\)}
\newcommand{\epz}{\(E_{p,{\rm z}}\)}
\newcommand{\rband}{r\(^\prime\)}
\newcommand{\gband}{g\(^\prime\)}
\newcommand{\iband}{i\(^\prime\)}

\begin{document}

\title{Long \thisgrb: an Off-axis Powerlaw Jet with Thermal Cocoon}

\author[0009-0002-7897-6110]{Utkarsh Pathak}
\email{utkarshpathak.07@gmail.com}
\affiliation{Department of Physics, Indian Institute of Technology Bombay, Powai, Mumbai 400076, India}

\author[0009-0002-8110-0515]{Sameer K. Patil}
\email{sameerkpatil@gmail.com}
\affiliation{Department of Physics, Indian Institute of Technology Bombay, Powai, Mumbai 400076, India}

\author[0009-0008-6644-5412]{Hitesh Tanenia}
\email{hitesh.tanenia@iitb.ac.in}
\affiliation{Department of Physics, Indian Institute of Technology Bombay, Powai, Mumbai 400076, India}

\author[0009-0001-4683-388X]{Tanishk Mohan}
\email{tanishk.mohan@iitb.ac.in}
\affiliation{Department of Physics, Indian Institute of Technology Bombay, Powai, Mumbai 400076, India}

\author[0000-0002-9364-5419]{Xander J. Hall}
\email{xjh@andrew.cmu.edu}
\affiliation{McWilliams Center for Cosmology and Astrophysics, Department of Physics, Carnegie Mellon University, 5000 Forbes Avenue, Pittsburgh, PA 15213, USA}

\author[0000-0002-5890-9298]{Yogesh Wagh}
\email{ywagh4430@gmail.com}
\affiliation{Department of Physics, Indian Institute of Technology Bombay, Powai, Mumbai 400076, India}

\author[0000-0002-7942-8477]{Viswajeet Swain}
\email{vishwajeet.s@iitb.ac.in}
\affiliation{Department of Physics, Indian Institute of Technology Bombay, Powai, Mumbai 400076, India}

\author[0009-0005-2987-0688]{Aditya Pawan Saikia}
\email{adityaps@iitb.ac.in}
\affiliation{Department of Physics, Indian Institute of Technology Bombay, Powai, Mumbai 400076, India}

\author[0000-0002-6112-7609]{Varun Bhalerao}
\email{varunb@iitb.ac.in}
\affiliation{Department of Physics, Indian Institute of Technology Bombay, Powai, Mumbai 400076, India}

 
\author[0000-0002-2184-6430]{Tomas Ahumada}
\email{tomas.ahumada@noirlab.edu}
\affiliation{Cerro Tololo Inter-American Observatory/NSF NOIRLab, Casilla 603, La Serena, Chile}

\author[0000-0003-3533-7183]{G. C. Anupama}
\email{gca@iiap.res.in}
\affiliation{Indian Institute of Astrophysics, II Block Koramangala, Bengaluru 560034, India}

\author[0000-0002-3927-5402]{Sudhanshu Barway}
\email{sudhanshu.barway@iiap.res.in}
\affiliation{Indian Institute of Astrophysics, II Block Koramangala, Bengaluru 560034, India}

\author[0009-0001-0574-2332]{Malte Busmann}
\email{m.busmann@physik.lmu.de}
\affiliation{University Observatory, Faculty of Physics, Ludwig-Maximilians-Universität, Scheinerstr.\ 1, 81679 Munich, Germany}
\affiliation{Excellence Cluster ORIGINS, Boltzmannstr.\ 2, 85748 Garching, Germany}

\author[0000-0002-8262-2924]{Michael W. Coughlin}
\email{cough052@umn.edu}
\affiliation{School of Physics and Astronomy, University of Minnesota, Minneapolis, MN 55455, USA}

\author[0000-0002-3168-0139]{Matthew J. Graham}
\email{mjg@caltech.edu}
\affiliation{Division of Physics, Maths and Astronomy, California Institute of Technology, 1200 E. California Blvd, Pasadena, CA 91125, USA}

\author[]{Daniel Gruen}
\email{daniel.gruen@lmu.de}
\affiliation{University Observatory, Faculty of Physics, Ludwig-Maximilians-Universität, Scheinerstr.\ 1, 81679 Munich, Germany}
\affiliation{Excellence Cluster ORIGINS, Boltzmannstr.\ 2, 85748 Garching, Germany}

\author[0000-0002-5936-1156]{Assaf Horesh}
\email{assafh@mail.huji.ac.il}
\affiliation{Racah Institute of Physics, The Hebrew University of Jerusalem, Jerusalem 91904, Israel}

\author[0000-0002-5619-4938]{Mansi M. Kasliwal}
\email{mansi@astro.caltech.edu}
\affiliation{Division of Physics, Mathematics, and Astronomy, California Institute of Technology, Pasadena, CA 91125, USA}

\author[0000-0003-2451-5482]{Russ R. Laher}
\email{laher@ipac.caltech.edu}
\affiliation{IPAC, California Institute of Technology, 1200 E. California Blvd, Pasadena, CA 91125, USA}

\author[0000-0002-8532-9395]{Frank J. Masci}
\email{fmasci@ipac.caltech.edu}
\affiliation{IPAC, California Institute of Technology, 1200 E. California Blvd, Pasadena, CA 91125, USA}

\author[0000-0002-6011-0530]{Antonella Palmese}
\email{apalmese@andrew.cmu.edu}
\affiliation{McWilliams Center for Cosmology and Astrophysics, Department of Physics, Carnegie Mellon University, 5000 Forbes Avenue, Pittsburgh, PA 15213, USA}

\author[0000-0003-1227-3738]{Josiah Purdum}
\email{jpurdum@caltech.edu}
\affiliation{Caltech Optical Observatories, California Institute of Technology, 1200 E. California Boulevard, Pasadena, CA 91125, USA}

\author[0000-0001-7357-0889]{Argyro Sasli}
\email{asasli@umn.edu}
\affiliation{School of Physics and Astronomy, University of Minnesota, Minneapolis, MN 55455, USA}
\affiliation{NSF Institute on Accelerated AI Algorithms for Data-Driven Discovery (A3D3)}

\author[0000-0001-7062-9726]{Roger Smith}
\email{rsmith@astro.caltech.edu}
\affiliation{Caltech Optical Observatories, California Institute of Technology, 1200 E. California Boulevard, Pasadena, CA 91125, USA}

\author[0000-0002-0772-6280]{Xiaoxiong Zuo}
\email{Xiaoxiong.Zuo@physik.lmu.de}
\affiliation{University Observatory, Faculty of Physics, Ludwig-Maximilians-Universität, Scheinerstr.\ 1, 81679 Munich, Germany}


\begin{abstract}
Some gamma-ray bursts (GRBs) exhibit precursor emission episodes preceding the main emission, 
with a quiescent period in between. The properties of the precursor emission and the duration of the quiescent interval are related to the central engine activity and jet formation processes, thus providing insights into the physics of GRBs.

We present a comprehensive analysis of the prompt emission and multi-wavelength afterglow of \thisgrb. Using detailed afterglow modeling, we find that the broadband data are best described by a powerlaw structured jet with a relatively narrow core (\(\theta_c \approx 0.8\degr\)), viewed moderately off-axis at a viewing angle \(\theta_v \approx 2.7\degr\). The isotropic-equivalent kinetic energy of the jet (\ekiso\ \(\approx 2.4 \times 10^{54}\)~erg) is on the higher side for typical GRBs. The precursor emission is well described by a blackbody spectrum with a temperature of \(kT \approx 13.2\)~keV and is separated from the main emission by a long quiescent interval of \(150\)~s. 

Put together, our results indicate that the precursor is likely to be a shock breakout from a cocoon formed by the interaction of the relativistic jet with the progenitor star. The resulting cocoon pressure and shock collimation naturally lead to the launch of a narrowly collimated jet, consistent with the jet geometry inferred from afterglow observations. The long quiescent interval may imply the central engine turn-off in addition to the effect of the off-axis geometry.

\end{abstract}



\section{Introduction}\label{sec:introduction}
Gamma-Ray Bursts (GRBs) are one of the most luminous transients in the universe, releasing up to \(10^{54}\)~erg of energy over timescales ranging from a fraction of a second to several minutes~\citep{Piran2004, KumarZhang2015}. Such bursts have been observed and characterized by numerous satellites over the past several decades, resulting in significant progress in our understanding of the physical properties of GRBs. Traditionally, GRBs are classified based on the duration (\tnt) of the \(90\%\) of prompt fluence, leading to two populations. Bursts with \tnt\ $< 2$~s are classified as short GRBs, which are linked to compact object mergers involving neutron stars, accompanied by kilonova emission~\citep{Eichler1989, Abbott2017GW170817}. In contrast, those with \tnt\ $> 2$~s are long GRBs~\citep{Kouveliotou1993}, which are associated with the core collapse of massive stars and are frequently accompanied by broad-lined Type~Ic supernovae~\citep{WoosleyBloom2006,modjaz2016}.

Some GRBs exhibit a precursor emission episode before the main emission in the prompt. These precursors are typically weaker, harder to detect, spectrally softer, and often are separated from the main emission by a quiescent interval ranging from a few seconds to a few minutes~\citep{Koshut1995,lazzati2005}. Several systematic searches for the precursors have been conducted in different missions: \emph{BATSE}~\citep{Koshut1995}, \swift~\citep{burlon2008} and \fermi/GBM~\citep{coppin2020}, the fraction of GRBs that show precursors depends upon the definition, energy range, sensitivity, and method of identification, giving a range of \(3\% - 20\%\) of all GRBs~\citep{coppin2020}.

The physical origin of GRB precursors remains an open question, with several models proposed to explain both their spectral diversity and the temporal separation between them and the main emission. One class of models attributes precursors to the same physical process that powers the main emission, where a temporary suppression of activity---caused by a change in the accretion rate onto the compact object---leads to a quiescent interval between the precursor and the main emission, without invoking a distinct emission mechanism~\citep{nappo2014, razzaque2003}. In a similar framework, intermittent central engine activity driven by fallback accretion can produce multiple emission episodes~\citep{wang2007}. In another scenario, the star initially collapses to form a proto-neutron star that launches a weak or short-lived jet, giving rise to the precursor emission. Subsequently, it collapses into a black hole that relaunches a relativistic jet responsible for the main emission~\citep{Metzger2011}. Alternatively, precursors may arise from a radiative transition in the outflow itself, as the ejecta evolves from an optically thick to an optically thin regime, producing a thermal precursor from photospheric emission before the onset of nonthermal radiation~\citep {2017SSRv..207...87B, Daigne2002, Hascoet2013}. A physically distinct interpretation associates GRB precursors with shock breakout from a cocoon formed as the relativistic jet propagates through the progenitor star. In this picture, energy deposited by the jet inflates a hot, mildly relativistic cocoon that eventually breaks out of the stellar surface, releasing quasithermal emission that is less collimated and more isotropic than the main jet emission~\citep{lazzati2005, ramirez2002}. Such cocoon shock-breakout emission can naturally account for thermal spectra, modest isotropic energies, and long quiescent intervals before the emergence of the highly collimated main jet.

The prompt emission of the burst is followed by long-lived multi-wavelength afterglow emission, produced as the relativistic ejecta decelerate in the surrounding circumburst medium~\citep{MeszarosRees1997, Sari1998}. Such emission, well described by a broadband synchrotron, can be fitted with a powerlaw~\citep{Sari1998}. An extensive follow-up of afterglow in different bands helps constrain several jet properties, including the structure and geometry of the jet. 

Early models of GRB jets often adopted a uniform ``top-hat'' configuration, characterized by sharp edges and a constant energy per unit solid angle. Subsequent theoretical developments and accumulating observational evidence increasingly support jets with angular structure, in which both the energy and the Lorentz factor are angle-dependent~\citep{rossi2002, kumar2003}. Such structured jets can arise naturally from jet propagation through dense environments and are supported by hydrodynamic simulations~\citep{lazzati2018, gottlieb2021}.

On-axis jets show a jet-break feature when bulk Lorentz factor decreases to \(\sim \theta_j^{-1}\)~\citep{sari1999}. In an alternative interpretation, the time of jet-break depends on the viewing angle \(\theta_v\) instead, in the case of the off-axis afterglows~\citep{rossi2002}. The combined effect of jet structure and viewing angle can lead to different features in the observed light curve: a rise followed by decay, or a plateau followed by decay, or decays at different slopes.

An interesting question is to link precursor activity with the angular structure and collimation of the jet, as the jet structure is expected to be impacted by jet-cocoon interaction~\citep{lazzati2005}. \thisgrb\ presents a compelling case study for exploring the connection between precursor emission, jet structure, and viewing geometry. The burst exhibits a clear thermal precursor preceding the main prompt emission, followed by a well-sampled afterglow. In this work, we analyze the prompt and afterglow properties of \thisgrb\ to investigate the physical origin of its precursor and to constrain the geometry and angular structure of its relativistic outflow.

\begin{figure*}[hbt!]
    \centering
    \includegraphics[width=1\linewidth]{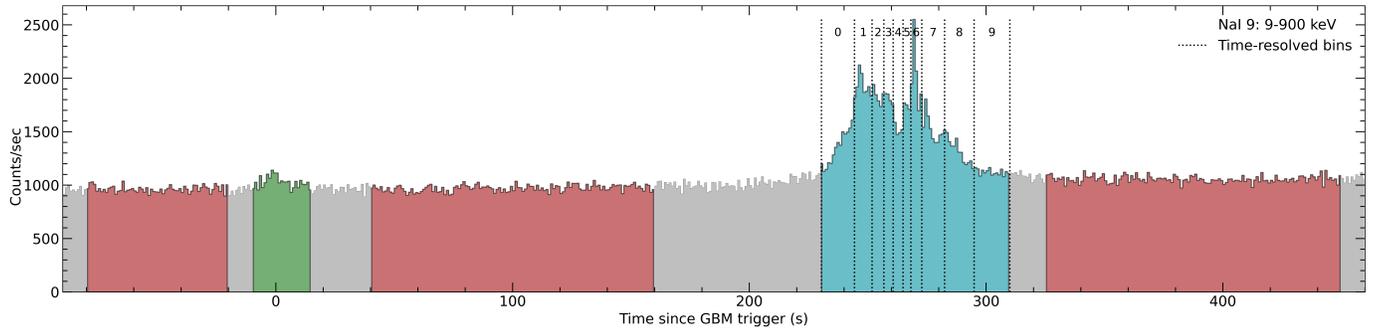}
    \caption{The light curve from the brightest NaI detector, \(n_9\), in the \(9-900\)~keV energy range. The green-shaded region marks the precursor emission episode, the cyan-shaded region marks the main emission episode. The red bins represent the background intervals considered. The vertical dashed line shows bins in the main emission episode used for time-resolved analysis. \label{fig:prompt_light_curve}}
\end{figure*}

This paper is organized as follows. Section~\ref{sec:observations} describes the observations of \thisgrb. In Section~\ref{sec:prompt}, we describe the prompt emission properties, including the precursor emission episode and main emission episode. Section~\ref{sec:afterglow-analysis} presents the multi-wavelength afterglow observations and simplistic analytical fitting done to determine the nature of the afterglow. Detailed afterglow modeling is done in Section~\ref{sec:afterglow_model}. In Section~\ref{sec:discussion}, we discuss the nature of the precursor, the inferred properties of the jet, and their relationship to each other.

\section{Observations and Data}\label{sec:observations}

\thisgrb\ was first detected and localized by \fermi/GBM~\citep{2025GCN.41839} with trigger time of precursor $T_0=$\,2025-09-16T13:29:21.01\,UT. The GBM lightcurve shows a precursor at $T_0$~s, which lasts for \(\sim\)25~s, leading to quiescence for 150~s and followed by the onset of main emission at $T_0+230$~s. The GBM team reports a $T_{90}\sim80$~s for the main emission in \(50-300\)~keV band~\citep{2025GCN.41876}. \astrosat/CZTI also detected the main emission and reported $T_{90}=48$~s in \(20-200\)~keV, $T_{90}=50$~s in \(100-500\)~keV energy range~\citep{2025GCN.41843}. Subsequently, the GRB was reported in several spacecraft/instrument circulars, including \emph{NuSTAR} detection of prompt emission~\citep{2025GCN.41871}, and \emph{CALET}/GBM~\citep{2025GCN.1442064382.CALET} and the Cadmium Zinc TElluride Radiation Imager (TERI) on board the ISS~\citep{2025GCN.42178}.

The optical counterpart for \thisgrb\ was first detected by Gravitational-wave Optical Transient Observer (GOTO) 10.43~hours after trigger~\citep{2025GCN.41847} with AB magnitude $18.90\pm0.11$ in the L band. \swift/XRT later imaged the GOTO position and detected the source at RA (J2000): $01^h 46^m 16.99^s$ and Dec (J2000): $36\degr 09\arcmin 54.70\arcsec$ with an error radius of 3.7~arcsec~\citep{2025GCN.41862}. Subsequently, an optical observation was reported by the GROWTH-India Telescope \citep{2025GCN.41858} at $T_0+28.13$ hours with AB magnitude $19.73\pm0.06$ in the \rband\ band. \cite{2025GCN.41863} reported a redshift of $2.015$. Using the \texttt{Plank2018} cosmology~\citep{planck2018} in \sw{astropy}~\citep{2022ApJ...935..167A} with $H_0 = 67.66~\mathrm{km~s^{-1}~Mpc^{-1}}$, $\Omega_m = 0.30966$, and $\Omega_\Lambda = 0.68884$, we calculate a luminosity distance of $16.07$~Gpc.

For broadband follow-up observations, we triggered several telescopes:
GROWTH-India Telescope (GIT), Himalayan Chandra
Telescope (HCT), Zwicky Transient Facility (ZTF), Spectral Energy Distribution Machine (SEDM), Acminute Microkelvin Imager (AMI), Fraunhofer Telescope at Wendelstein Observatory (FTW), as part of the GROWTH Collaboration~\citep{mansi2019}. We use the ZTF Fritz marshal to trigger follow-up observations and to archive the photometric data~\citep{2019JOSS....4.1247V, Coughlin_2023}. The observations and data reduction are described in the Appendix~\ref{appendix:afterglow}. We also used selective public data from various circulars reported on the General Coordinate Network (GCN). All the data used in this work and their corresponding sources are
listed in Table \ref{tab:afterglow_obs_table} in the appendix. 


\section{Prompt emission: analysis}\label{sec:prompt}
We present results of the prompt emission analysis, divided into the main emission episode (Section~\ref{sec:main_emission}) and the precursor episode (Section~\ref{sec:precursor}). Implications of the inferred properties are in the discussion (Section~\ref{sec:discussion}).

\subsection{Main Emission Episode}\label{sec:main_emission}
In Section~\ref{sec:TI_prompt}, we discuss time-integrated analysis to understand the broad characteristics of the main emission episode, followed by time-resolved analysis to check for the evolution of spectral parameters in Section~\ref{sec:TR_prompt}.

\begin{deluxetable*}{ccccccc}[hbt!]
\tablecaption{Best fit spectral parameters for the time-resolved analysis of \thisgrb\ using Band function for main emission. Analysis described in Section~\ref{sec:TR_prompt}.\label{tab:band_time_resolved}}
\tablehead{
\colhead{Bin} &
\colhead{$T_{\mathrm{start}}$ (s)} &
\colhead{$T_{\mathrm{stop}}$ (s)} &
\colhead{$K$} &
\colhead{$\alpha$} &
\colhead{$E_p$ (keV)} &
\colhead{BIC}
}
\startdata
0 & 230 & 244 & $0.01 \pm 0.002$ & $-0.88 \pm 0.09$ & $257.82 \pm 39.13$ & 2857 \\
1 & 244 & 252 & $0.04 \pm 0.004$ & $-0.88 \pm 0.06$ & $193.10 \pm 17.70$ & 2499 \\
2 & 252 & 257 & $0.04 \pm 0.01$ & $-0.87 \pm 0.09$ & $160.74 \pm 20.30$ & 2233 \\
3 & 257 & 261 & $0.04 \pm 0.01$ & $-0.99 \pm 0.12$ & $103.85 \pm 14.04$ & 2067 \\
4 & 261 & 265 & $0.06 \pm 0.02$ & $-0.57 \pm 0.22$ & $90.90 \pm 13.93$ & 2030 \\
5 & 265 & 268 & $0.05 \pm 0.02$ & $-0.81 \pm 0.16$ & $106.84 \pm 17.36$ & 1897 \\
6 & 268 & 273 & $0.06 \pm 0.01$ & $-0.96 \pm 0.11$ & $101.11 \pm 12.50$ & 2196 \\
7 & 273 & 282 & $0.02 \pm 0.01$ & $-1.25 \pm 0.14$ & $62.16 \pm 9.39$ & 2635 \\
8 & 282 & 295 & $0.03 \pm 0.02$ & $-0.85 \pm 0.26$ & $65.75 \pm 12.52$ & 2728 \\
9 & 295 & 310 & $0.01 \pm 0.01$ & $-0.76 \pm 0.51$ & $76.74 \pm 20.50$ & 2883 \\
\enddata
\end{deluxetable*}

\subsubsection{Time-Integrated Analysis}\label{sec:TI_prompt}
We performed time-integrated spectral analysis of the main emission episode using NaI and BGO detectors of \fermi\ Gamma-ray Burst Monitor~\citep[\fermi/GBM;][]{fermigbm}. For our analysis, we used detectors $n_a$ and $n_9$, along with the BGO detector $b_1$, which have viewing angles relative to the source $\le 60 \degr$. The source interval considered was from $T_{0}+230$~s to $T_{0}+310$~s. We fitted the background using three intervals: from $T_{0}-80$~s to $T_{0}-20$~s, $T_{0}+40$~s to $T_{0}+160$~s, and from $T_{0}+325$~s to $T_{0}+450$~s (see Figure~\ref{fig:prompt_light_curve}). We used the energy range of $9-900$~keV for NaI and $250-30000$~keV for BGO. Using the Multi-Mission Maximum Likelihood Framework~\citep[\sw{3ML};][]{2015arXiv150708343V}, spectra were extracted for each detector within these ranges.

We then modeled the spectra using typical models, the standard Band model, powerlaw, cutoff powerlaw, blackbody, and powerlaw $+$ blackbody. The Band function provided the best fit to the data, yielding the lowest Bayesian Information Criteria (BIC) value and effectively constraining all parameters. From the Band model~\citep{band1993}, we obtained a low-energy index \(\alpha = -1.04^{+0.06}_{-0.04}\), a high-energy index \(\beta = -2.00^{+0.03}_{-0.04}\), and an observed peak energy \(\ E_p = 140^{+11}_{-13}\)~keV. From the spectral analysis, we further estimated the fluence as $(6.6 \pm 0.2) \times 10^{-5}$~\eg, and the isotropic-equivalent energy to be \eiso$= (6.8 \pm 0.2) \times 10^{53}$~erg.

As discussed in Section~\ref{sec:global_relations}, these values are consistent with typical long GRBs.

\subsubsection{Time-Resolved Analysis}\label{sec:TR_prompt}

\begin{figure}[hbt!]
    \centering
    \includegraphics[width=0.45\textwidth]{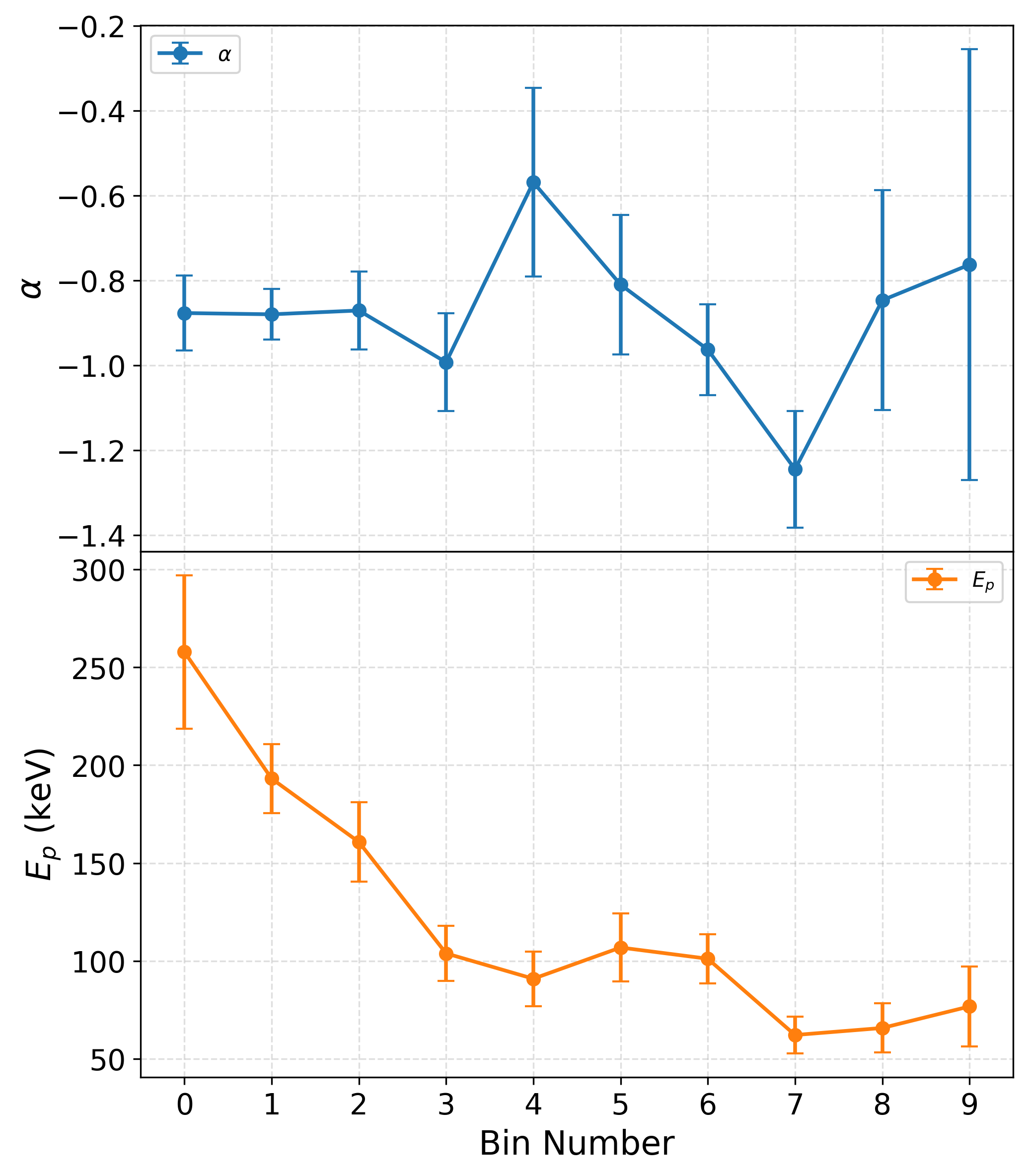}
    \caption{Time evolution of the parameters of the best-fit Band function for the main emission episode. \(E_p\) shows decline across the main emission episode, \(\alpha
    \) is largely consistent with no evolution.}
    \label{fig:main_emission_TR}
\end{figure}

We performed time-resolved spectral analysis of the main emission episode using the same set of NaI and BGO detectors as Section~\ref{sec:TI_prompt}. The source and background interval were also kept the same as the time-integrated analysis.

For the time-resolved analysis, we first defined intervals (bins) using the Bayesian Blocks algorithm for the brightest detector $n_9$, which produced 25 segments. We then evaluated the photon counts and their statistical significance for each bin using the built-in \texttt{3ML} functions. Some of the intervals were very narrow, with low counts, and had unconstrained fits to spectra. To ensure adequate statistics, we merged bins to get a subset of 10 intervals (Figure~\ref{fig:main_emission_TR}). Spectra were extracted for each of these bins in the energy range of $9-900$~keV for the NaI detectors and $250-30000$~keV for the BGO. For each of these intervals, we examined the spectra from each detectors to identify the effective energy ranges where the background did not dominate and established the final energy ranges for further analysis to be $9-290$~keV for $n_{0}$, $9-330$~keV for $n_{9}$, and $250-4900$~keV for $b_{1}$. For the fits, we adopted the standard Band function as our baseline spectral model. We fixed the \(\beta\) to the time-integrated value of \(-2\) as it was not constrained, and it was consistent with no evolution across the bins. The results (Table~\ref{tab:band_time_resolved}) are discussed further in Section~\ref{sec:nature_main}. 


\subsection{Precursor Emission Episode}\label{sec:precursor}

\begin{table}[hbt!]
\centering
\caption{Comparing various models for precursor emission episode using BIC. From \(\Delta\)BIC, we conclude the blackbody to be the best fit model.}
\begin{tabular}{lcc}
\hline
\hline
Model & BIC & $\Delta$BIC\\
\hline
Blackbody & 2393 & 0 \\
Power Law & 4324 & 1931 \\
Band & 3716 & 1323 \\
Cutoff Power Law & 4299 & 1906 \\
Power Law + Blackbody & 4005 & 1612 \\
Cutoff Power Law + Blackbody & 4072 & 1679 \\
\hline
\end{tabular}
\label{tab:precursor_stats}
\end{table}

\begin{figure*}[htb!]
    \centering
    \includegraphics[width=0.49\textwidth]{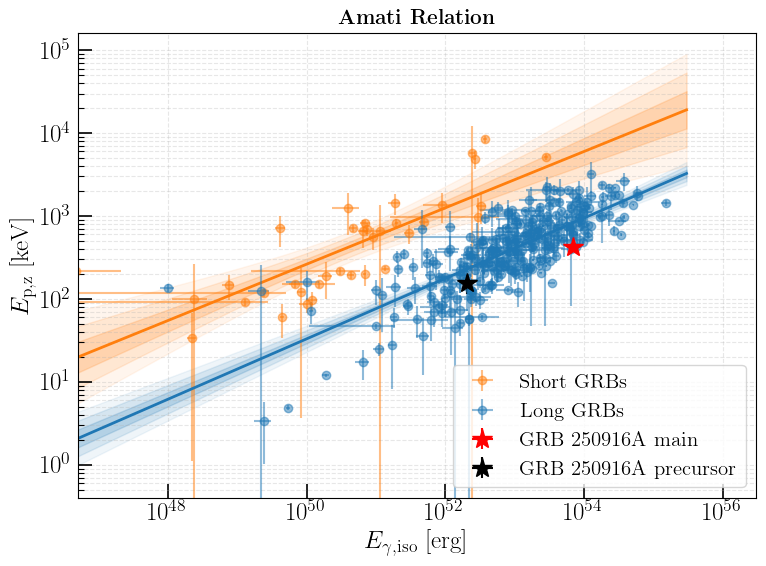}
    \includegraphics[width=0.49\textwidth]{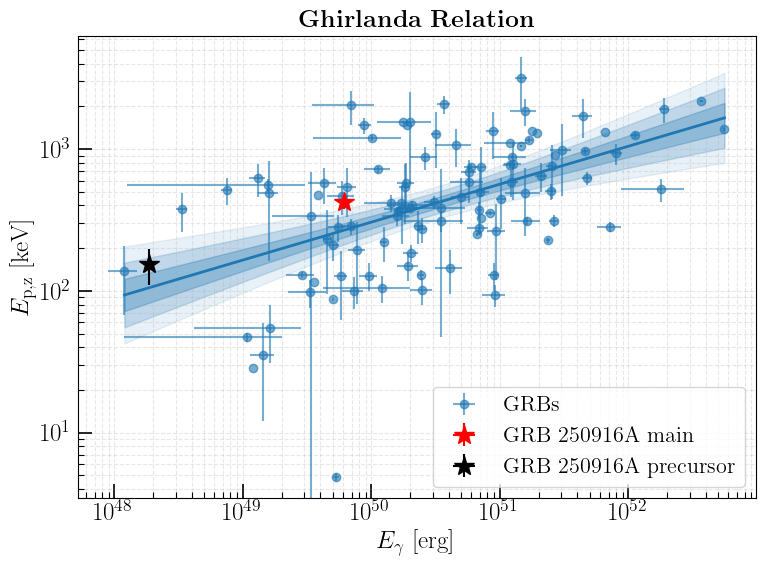}
    \includegraphics[width=0.49\textwidth]{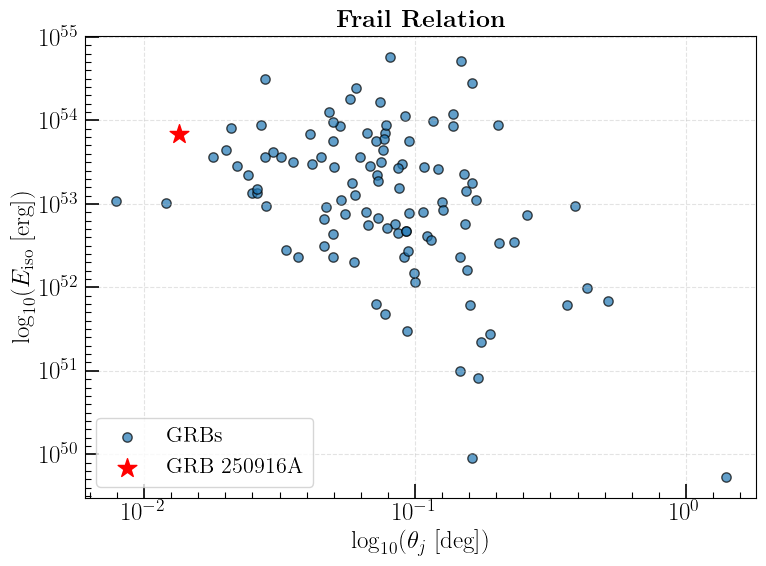}
    \includegraphics[width=0.49\textwidth]{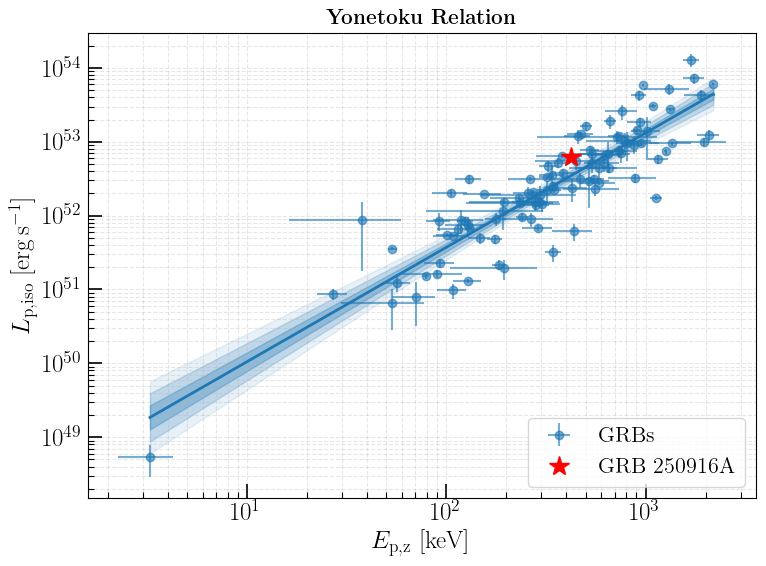}
    \caption{GRB global relations based on Band model fit to precursor and main emission episodes. Top left: Amati relation, \epz\ vs \egmiso. Top right: Ghirlanda relation, \epz\ vs \(E_\gamma\). Bottom left: Frail, \egmiso\ vs \(\theta_{j}\). Bottom right: Yonetoku relation, \epz\ vs - \(L_{p,{\rm iso}}\)}
    \label{fig:amati_ghirlanda_frail_yonetoku}
\end{figure*}

We undertook time-integrated spectral analysis of the precursor emission episode using \fermi/GBM data. We excluded the BGO detectors as they were entirely background-dominated and used \(n_a\) and \(n_9\) NaI detectors. We kept the same background intervals as given in Section~\ref{sec:TI_prompt}. The source interval was selected to be from $T_{0}-10$~s to $T_{0}+15$~s (Figure~\ref{fig:prompt_light_curve}). Using \texttt{3ML}, we generated spectra for all the chosen detectors. The energy range for NaI detectors was restricted to \(10-400\)~keV; beyond this range, the background dominated. The spectra were then modeled using the same set of functions mentioned in Section~\ref{sec:TI_prompt}. Model comparison was performed using the Bayesian Information Criterion (BIC;~\citealt{1978AnSta...6..461S,2007MNRAS.377L..74L}). We quantify the model preference with $\Delta\mathrm{BIC}$ and values of $\Delta{\rm BIC} > 10$ are considered decisive in favor of a model~\citep{KassRaftery1995, 2007MNRAS.377L..74L}. Based on the \(\Delta\)BIC values of the models discussed above (Table~\ref{tab:precursor_stats}) we found that a single-component blackbody provides the best fit to the data, giving the fluence of precursor to be \((9.4 \pm 0.9)\times 10^{-7}\)~\eg, and the corresponding \eiso\(= (9.6 \pm 0.9) \times 10^{51}\)~erg. The blackbody has temperature of \(kT = 13.2_{-2.1}^{+1.5}\)~keV.



\subsection{GRB Global Relations}\label{sec:global_relations}

Global relations for GRBs conventionally use a Band spectral fit. While a blackbody gives a better fit to the precursor, here we use our best-fit Band values of \(\alpha_{\rm precursor} = 0.19^{+0.56}_{-0.68}\), \(E_p = 51^{+13}_{-6.9}\)~keV, i.e., \(E_{p,z} = 154^{+39}_{-21}\)~keV; we fix \(\beta_{\rm precursor} = -2\), obtained from time-integrated analysis of the main emission episode. We get the fluence of precursor to be \((2.0 \pm 0.2)\times 10^{-6}\)~\eg, and the corresponding \eiso\(= (2.1 \pm 0.2) \times 10^{52}\)~erg. For Ghirlanda and Frail relation, we have used the inferred jet half-opening from afterglow modeling and assumed same angle for precursor (Section~\ref{sec:afterglow_model}). As seen in Figure~\ref{fig:amati_ghirlanda_frail_yonetoku}, both the main episode and the precursor episode of \thisgrb\ are consistent with the Amati correlation~\citep{amati2006}, the Ghirlanda relation~\citep{ghirlanda2005}, and the Yonetoku relation~\citep{yonetoku2004}. \thisgrb\ also largely follows the Frail relation~\citep{frail2001}.


\begin{figure}[hbt!]
    \centering
    \includegraphics[width=\columnwidth]{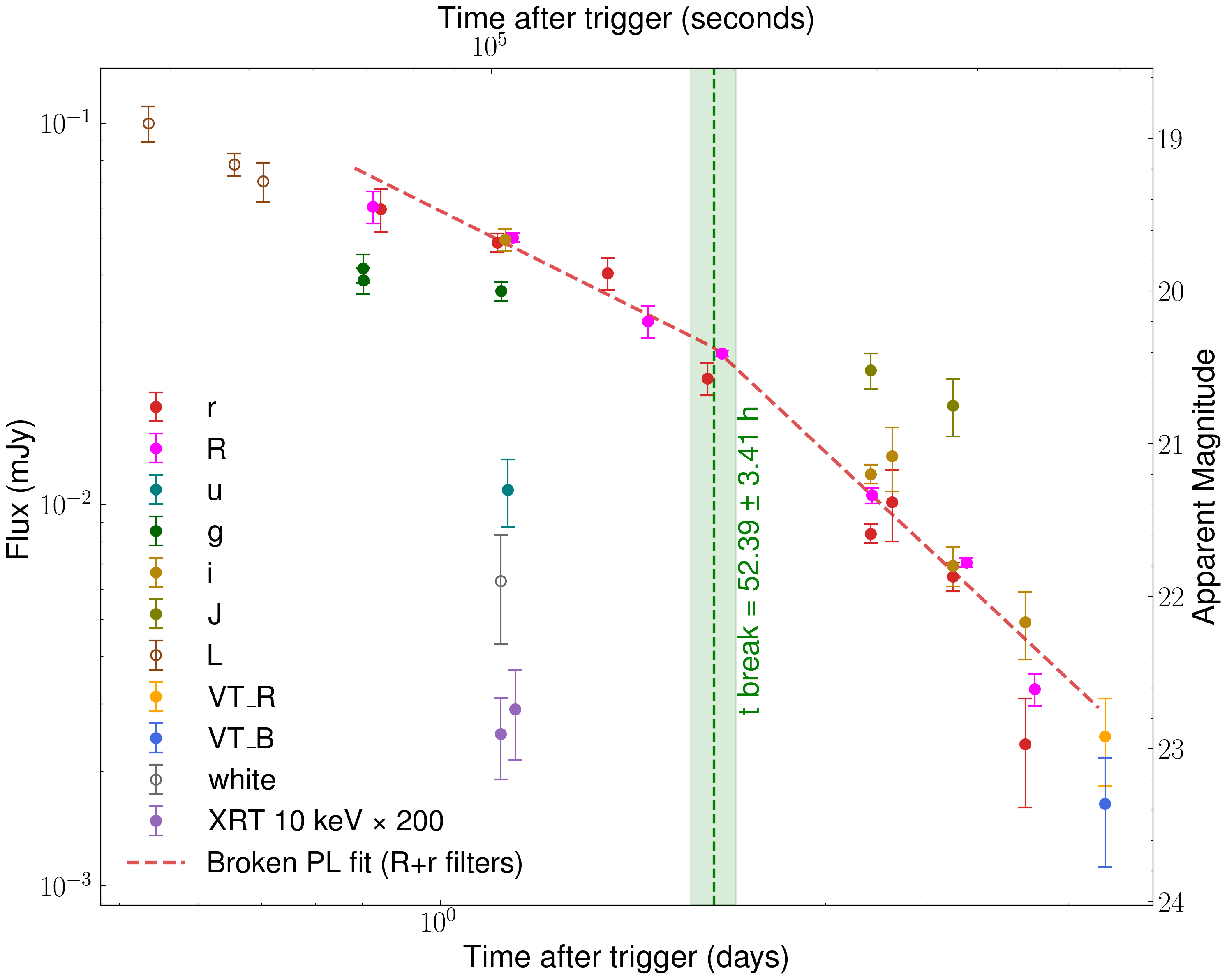}
    \caption{Multi-wavelength afterglow light curves of \thisgrb\ in X-ray (black), optical ($u$, $g$, $r$, $i$, $R$, $white$, $L$, $VT_R$, $VT_B$, and $J$ bands). The filled markers represent fluxes corrected for galactic extinction while the hollow markers represent uncorrected values. The light curves are described by a broken powerlaw with an initial $\alpha_1 = 1.06 \pm 0.10$ followed by a steep decay $\alpha_2 = 2.07 \pm 0.04$ after the break at $t_{\mathrm{break}} = 53 \pm 3$ hours (green vertical line).}
    \label{fig:multiband-light-curve}
\end{figure}

\section{Afterglow}\label{sec:afterglow-analysis}
As mentioned in Section~\ref{sec:introduction}, the afterglow consists of a broadband synchrotron emission that occurs when GRB ejecta interacts with the circumburst medium~\citep{1997ApJ...489L..37S, 2002ApJ...568..820G}. The observed afterglow emission can typically be described by a simple powerlaw, such that flux spectral density \(F_\nu \propto t^{-\alpha} \nu^{-\beta}\)~\citep{1997ApJ...489L..37S}. 

In the case of \thisgrb\ the optical afterglow lightcurve shows a clear break and cannot be well fit by a simple powerlaw. Instead, we adapt a broken powerlaw model (Equation~\ref{eq:bkn_pl}):
\begin{equation}
    F(t) =
\begin{cases}
F_{\mathrm{break}} \left( \dfrac{t}{t_{\mathrm{break}}} \right)^{-\alpha_1}, & t < t_{\mathrm{break}}, \\
F_{\mathrm{break}} \left( \dfrac{t}{t_{\mathrm{break}}} \right)^{-\alpha_2}, & t \geq t_{\mathrm{break}},
\label{eq:bkn_pl}
\end{cases}
\end{equation}
where \(F(t)\) is the flux expected at a given time \(t\) since the burst, \(F_{\rm{break}}\) is the normalization constant, \(t_{\rm break}\) is the time at of break in the powerlaw, \(\alpha_1, \alpha_2\) are indices before and after the break, respectively.

We use data from the GROWTH Collaboration, supplemented by select publicly-reported observations, presented in the appendix in Table~\ref{tab:afterglow_obs_table}. The optical afterglow exhibits a clear temporal break in its light curve (see Figure~\ref{fig:multiband-light-curve}). The afterglow is most densely sampled in the \rband\ and $R$ bands, and therefore further used in fitting the broken powerlaw model, which yields $\alpha_1 = 1.06 \pm 0.10$, $\alpha_2 = 2.07 \pm 0.04$, and $t_{\mathrm{break}} = 53 \pm 3$~hours. \\
Assuming a powerlaw dependence of the flux density on frequency, $F_\nu \propto \nu^{-\beta}$, we compute the spectral index $\beta$. Using near contemporaneous data at 29~hours since the burst, we measure a spectral index of $\beta_o = 1.38 \pm 0.14$ across $u,g,r,i$ bands before the temporal break in the lightcurve and $\beta_{oJ} = 1.21 \pm 0.32$ across $r,i,J$ bands post-break. We note that the spectral index for X-ray to be $\beta_x = 0.95 \pm 0.88$~\citep{evans2009}. The spectral index across the infrared, optical, and X-ray are consistent with each other, hinting towards slow cooling spectral regime.

\section{Afterglow Modeling}\label{sec:afterglow_model}

\begin{table*}[hbt!]
\centering
\caption{Posterior constraints for \thisgrb\ under different jet models. Fixed parameters: $\Gamma_0 = 426$, $A_V = 0$, $z=2.015$. BIC and $\Delta$BIC comparison for the three jet configurations fitted to the afterglow of \thisgrb.}
\label{tab:jet_model_fit}
\renewcommand{\arraystretch}{1.3}
\begin{tabular}{l c c c c}
\hline
\hline
& & \multicolumn{3}{c}{Posterior} \\
\cline{3-5}
Parameter
& Common Prior
& Tophat Jet
& Powerlaw Jet
& Gaussian Jet \\
\hline
$\log E_{0}$      
& $[\phantom{0}53.0,\;55.0]$
& $53.99^{+0.10}_{-0.15}$
& $54.38^{+0.14}_{-0.13}$
& $54.44^{+0.06}_{-0.12}$ \\

$\log \epsilon_B$
& $[-8.0,\;-1.0]$ 
& $-1.51^{+0.32}_{-0.40}$
& $-2.00^{+0.19}_{-0.21}$
& $-1.43^{+0.25}_{-0.25}$ \\

$\log \epsilon_e$
& $[-1.5,\;-0.8]$ 
& $-1.41^{+0.09}_{-0.06}$
& $-1.17^{+0.12}_{-0.12}$
& $-1.41^{+0.09}_{-0.05}$ \\

$\log n_0$
& $[-2.0,\;\phantom{0}0.0]$
& $-0.50^{+0.20}_{-0.19}$
& $-0.41^{+0.27}_{-0.25}$
& $-1.11^{+0.27}_{-0.21}$ \\

$\log \theta_{\rm c}$
& $[-3.0,\;-0.5]$ 
& $-1.15^{+0.02}_{-0.02}$
& $-1.87^{+0.08}_{-0.09}$
& $-1.83^{+0.05}_{-0.04}$ \\

$\log \theta_{\rm v}$
& $[-5.0,\;-0.5]$ 
& $-2.91^{+0.64}_{-0.94}$
& $-1.32^{+0.04}_{-0.03}$
& $-1.40^{+0.06}_{-0.03}$ \\

$p$
& $[\phantom{0}2.01,\;3.00]$
& $2.035^{+0.01}_{-0.01}$
& $2.23^{+0.05}_{-0.04}$
& $2.24^{+0.21}_{-0.06}$ \\

$s$
& $[\;\phantom{0}1.0,\;\;\phantom{0}8.0]$
& --
& $2.23^{+0.32}_{-0.22}$
& -- \\

BIC
& --
& 472
& 447
& 470 \\

$\Delta$BIC
& --
& 25
& 0
& 23 \\
\hline
\hline
\end{tabular}
\end{table*}

We model the broadband afterglow emission of \thisgrb\ using the 
\texttt{jetsimpy} forward-shock framework, fitting the optical and X-ray light curves simultaneously. \texttt{jetsimpy}~\citep{2024ApJS..273...17W} is a publicly available package that computes the synchrotron emission from relativistic jets interacting with an external medium. The code approximates the blast wave as a thin relativistic shell and self-consistently tracks lateral spreading, radiative cooling, and relativistic beaming effects. To explore the parameter space efficiently, we employed the nested sampling algorithm \textsc{MultiNest}~\citep{2009MNRAS.398.1601F} through its Python interface \texttt{PyMultiNest}~\citep{2016ascl.soft06005B} with 2000 live points. Three jet configurations are explored: a uniform tophat jet (Section~\ref{sec:tophat}), a powerlaw structured jet (Section~\ref{sec:powerlaw}), and a Gaussian structured jet (Section~\ref{sec:gaussian}). For model comparison, we compute the Bayesian Information Criterion (BIC) for each fit. 

The afterglow modeling requires an assumption for the initial bulk Lorentz factor, $\Gamma_0$, which governs the early deceleration of the relativistic outflow. However, for afterglow observations obtained well after the deceleration time, the model light curves are largely insensitive to the exact value of $\Gamma_0$, as the blast wave has already transitioned to the self-similar deceleration phase. Consequently, $\Gamma_0$ is poorly constrained by the data and remains strongly degenerate with other model parameters. Motivated by population studies of long GRBs in a uniform interstellar medium, which find a mean initial Lorentz factor of $\Gamma_0 \simeq 426$~\citep{zhang2024}, we fix $\Gamma_0 = 426$ for all model fits. Tests performed with different fixed values of $\Gamma_0$ within the typical range for long GRBs do not significantly affect the inferred jet structure or viewing geometry. The adopted priors and best-fit parameters for each jet model are summarized in Table~\ref{tab:jet_model_fit}.

\begin{figure*}[hbt!]
\centering
\begin{subfigure}[b]{\linewidth}
    \centering
    \includegraphics[width = 0.7\linewidth]{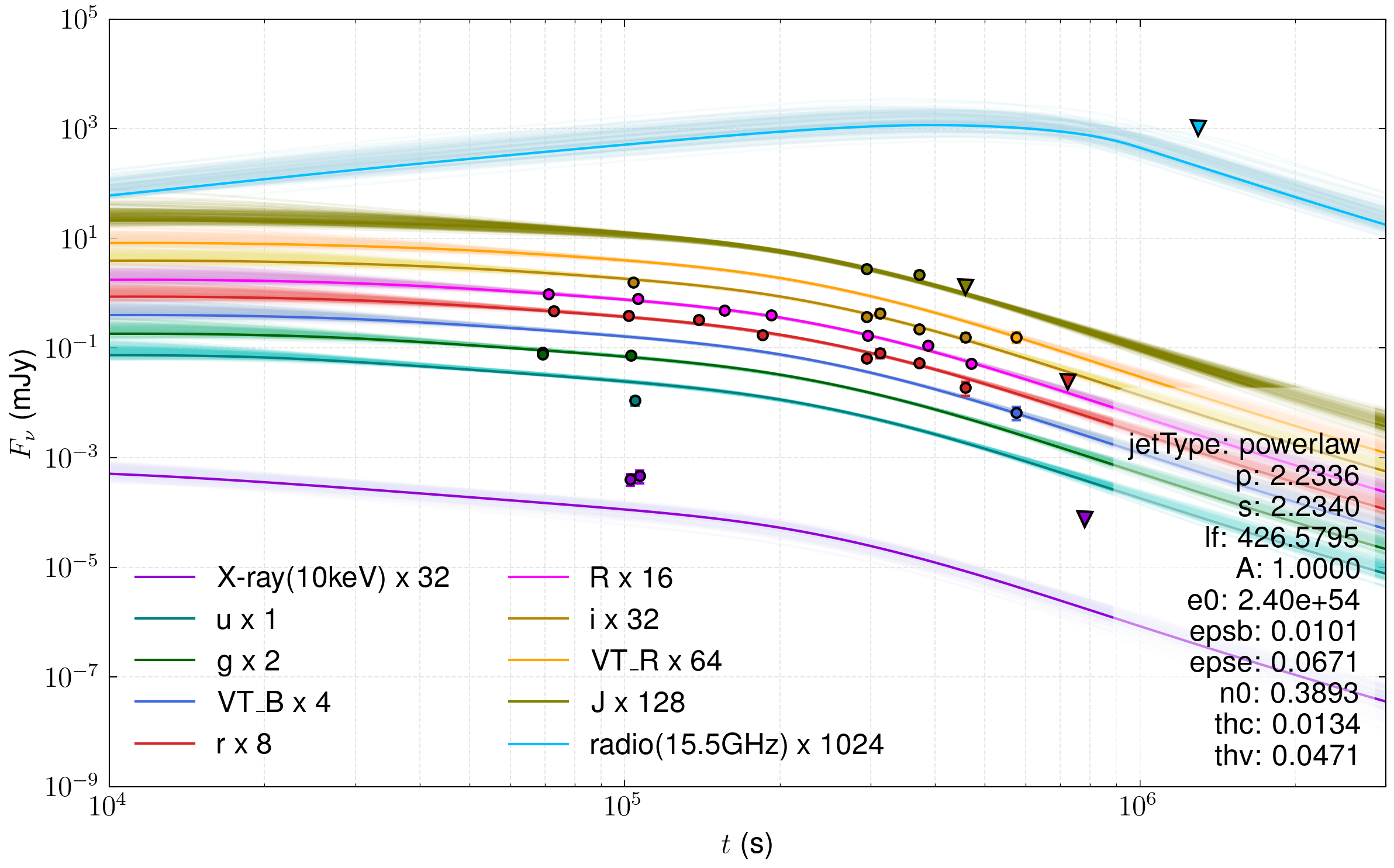}
    \label{fig:powerlaw_lc}
\end{subfigure}
\begin{subfigure}[b]{\linewidth}
    \centering
    \includegraphics[width = 0.68\linewidth]{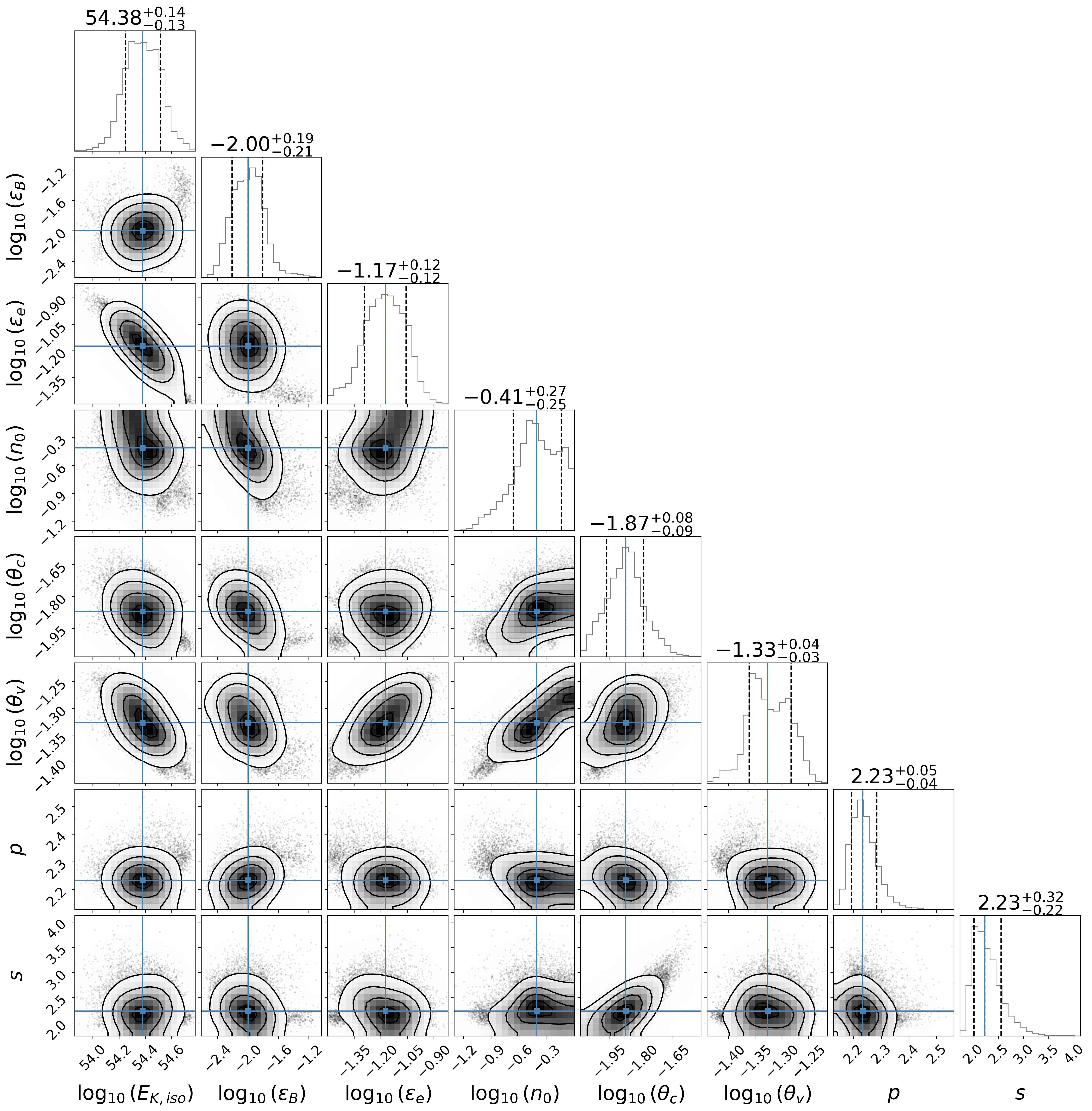}
    \label{fig:powerlaw_corner}
\end{subfigure}

\caption{Model light-curve and posterior constraints for \thisgrb\ with the best-fitting Powerlaw jet model. Top: \texttt{jetsimpy} afterglow predictions overlaid on the optical and X-ray observations, downward triangles are upper limits. Bottom: Posterior corner plots obtained with \textsc{MultiNest}.}
\label{fig:powerlaw_jet_model}
\end{figure*}

\subsection{Uniform Tophat Jet}
\label{sec:tophat}

The uniform tophat jet model assumes a constant energy per unit solid angle within a sharp half-opening angle $\theta_{\rm c}$, with no angular structure beyond it~\citep{1999ApJ...519L..17S}. The posterior distributions of parameters for this model show modest correlations and significant uncertainty in a few parameters (notably $n_0$ and $\epsilon_{\rm e}$). The predicted light curves reasonably reproduce the temporal evolution in both the optical and X-ray bands, including a clear jet-break, consistent with expectations for a narrow, uniform jet viewed on-axis. However, this fit is not statistically preferred over the other fits (Table~\ref{tab:jet_model_fit}). 

\subsection{Powerlaw Structured Jet}
\label{sec:powerlaw}

In the powerlaw structured jet model, the energy per unit solid angle and Lorentz factor decrease with angle from the jet axis as:
\begin{equation}
E(\theta) = E_{k,{\rm iso}} \left[ 1 + \left( \frac{\theta}{\theta_c} \right)^{2} \right]^{-s/2}, \label{eq:powerlaw_jet}
\end{equation}

\begin{equation}
\Gamma(\theta) = (\Gamma_{0} - 1) \left[ 1 + \left( \frac{\theta}{\theta_c} \right)^{2} \right]^{-s/2} + 1, \label{eq:powerlaw_jet2}
\end{equation}

where \ekiso\ is the on-axis (core) isotropic-equivalent energy, $\theta$ is the angle from the jet axis, $\theta_c$ is the core angle of the jet, and $s$ is the powerlaw index controlling how steeply energy falls off. $\Gamma_0$ is the initial Lorentz factor~\citep{2024ApJS..273...17W}.  

The parameters are quite tightly constrained (Table~\ref{tab:jet_model_fit}) with no strong correlations in the posterior distribution (Figure~\ref{fig:powerlaw_jet_model}). The model reproduces the optical bands well but slightly under-predicts the X-ray flux and slightly over-predicts the u-band flux. The predicted temporal evolution exhibits a smooth transition between the early- and late-time decay phases, rather than a sharp jet break. An off-axis viewing geometry in a powerlaw structured jet can produce such behavior. The gradual emergence of emission from wider angles mimics the steepening of the light curve without a sharp break~\citep{paz2020, hendrik2013}. The inferred \ekiso\ of \(2.4_{-0.6}^{+0.9} \times 10^{54}\)~erg is on the higher side but well within the distribution of \ekiso\ for long GRBs~\citep{wang2018,aksulu2022}. While the inferred jet core angle of \(\theta_c = 0.8^{+0.2}_{-0.1}\)~degrees is on the narrower side; similar angles have been seen in other GRBs~\citep{boatlhaaso, 230812Bgrandma, ronning2020}. 

\subsection{Gaussian Structured Jet}
\label{sec:gaussian}

The Gaussian structured jet model assumes that the energy per unit solid angle follows
\begin{equation}
    E(\theta) = E_{k,{\rm iso}} \exp\!\left(-\frac{\theta^2}{2\theta_{\rm c}^2}\right),
\end{equation}
with a similar profile for the initial Lorentz factor~\citep[e.g.][]{2014PASA...31....8G}. We infer the seven fitted parameters in Table~\ref{tab:jet_model_fit}. The posterior distributions show reasonably tight constraints, though not as narrow as those obtained for the powerlaw jet. 

\subsection{Model Comparison}
\label{sec:model_comparison}

As discussed in Section~\ref{sec:precursor}, we are using BIC values for comparing different models. Based on the \(\Delta\)BIC values of the three models discussed above (Table~\ref{tab:jet_model_fit}), we conclude that the powerlaw structured jet model is strongly supported by observed data. Physically, the structured jet interpretation is compelling: a powerlaw jet viewed slightly off-axis can account for the relative smoothness of the jet break in both optical and X-ray bands.

\section{Discussion}\label{sec:discussion}

\subsection{Nature of the Main Emission}\label{sec:nature_main}

The main emission episode of \thisgrb\ has a duration of $\sim80$~s and is well described by a Band function. The time-integrated spectral parameters are a low-energy index $\alpha = -1.04^{+0.06}_{-0.04}$, a high-energy index $\beta = -2.00^{+0.03}_{-0.04}$, and an observed peak energy $E_p = 140^{+11}_{-13}$~keV. These values are consistent with the median spectral properties of long GRBs observed by \fermi/GBM over a decade~\citep{fermi10yr}, indicating that the prompt emission of \thisgrb\ is spectrally typical of the long-GRB population. The isotropic-equivalent energy of the main emission is \eiso$= (6.8 \pm 0.2)\times10^{53}$~erg, on the higher side of typical long GRBs.

Time-resolved spectral analysis reveals a monotonic decrease in $E_p$ from $\sim250$~keV to $\sim60$~keV over the duration of the burst, consistent with the canonical hard-to-soft spectral evolution commonly observed in long GRBs~\citep{ford1995}. Such evolution can be physically interpreted as the result of radiative cooling and adiabatic expansion of the emitting region, leading to a progressive decrease in the characteristic electron Lorentz factor and magnetic field strength~\citep{uhm2018}. 
As discussed in Section~\ref{sec:global_relations}, the main emission of \thisgrb\ is consistent with the Amati, Ghirlanda, and Yonetoku relations, further supporting its classification as a typical long GRB.

\subsection{Perspective of Observations}

The ratio of viewing angle to jet half-opening angle tells us how off we are from the jet axis. For \thisgrb\ we have \({\theta_v}/{\theta_c} \simeq 3.4\). The structure of the jet is powerlaw with the index of \(s = 2.23_{-0.22}^{+0.32}\), marginally higher than the expected canonical powerlaw slope \(s \approx 2\)~\citep{rossi2002, zhang2002}. This suggests a slightly stronger lateral energy gradient or aggressive jet collimation; jets with even higher values of \(s\) have been observed in some GRBs~\citep{powerlawcatalog, swain2025}. Since the first detection of the afterglow was \(\sim 10\)~hours post trigger, there are no data covering the early phase of the jet evolution where the observed lightcurve might have peaked. 
As discussed in Section~\ref{sec:introduction}, the break in the lightcurve of \thisgrb\ can also be explained by the viewing angle rather than a jet half-opening angle~\citep{rossi2002}.

\subsection{Long Quiescence Interval}

\cite{coppin2020} analyzed the quiescent intervals between precursor and main emission episodes in 209 \fermi-detected GRBs and found that the distribution is best described by a bimodal function. The short-gap component was interpreted as arising from photospheric emission, typically thermal in nature, while longer quiescent intervals were attributed to an effective suppression of the central engine, possibly driven by variations in the accretion rate. In the latter case, similar spectral characteristics between the precursor and main emission are seen.

\begin{figure}[hbt!]
    \centering
    \includegraphics[width=1\linewidth]{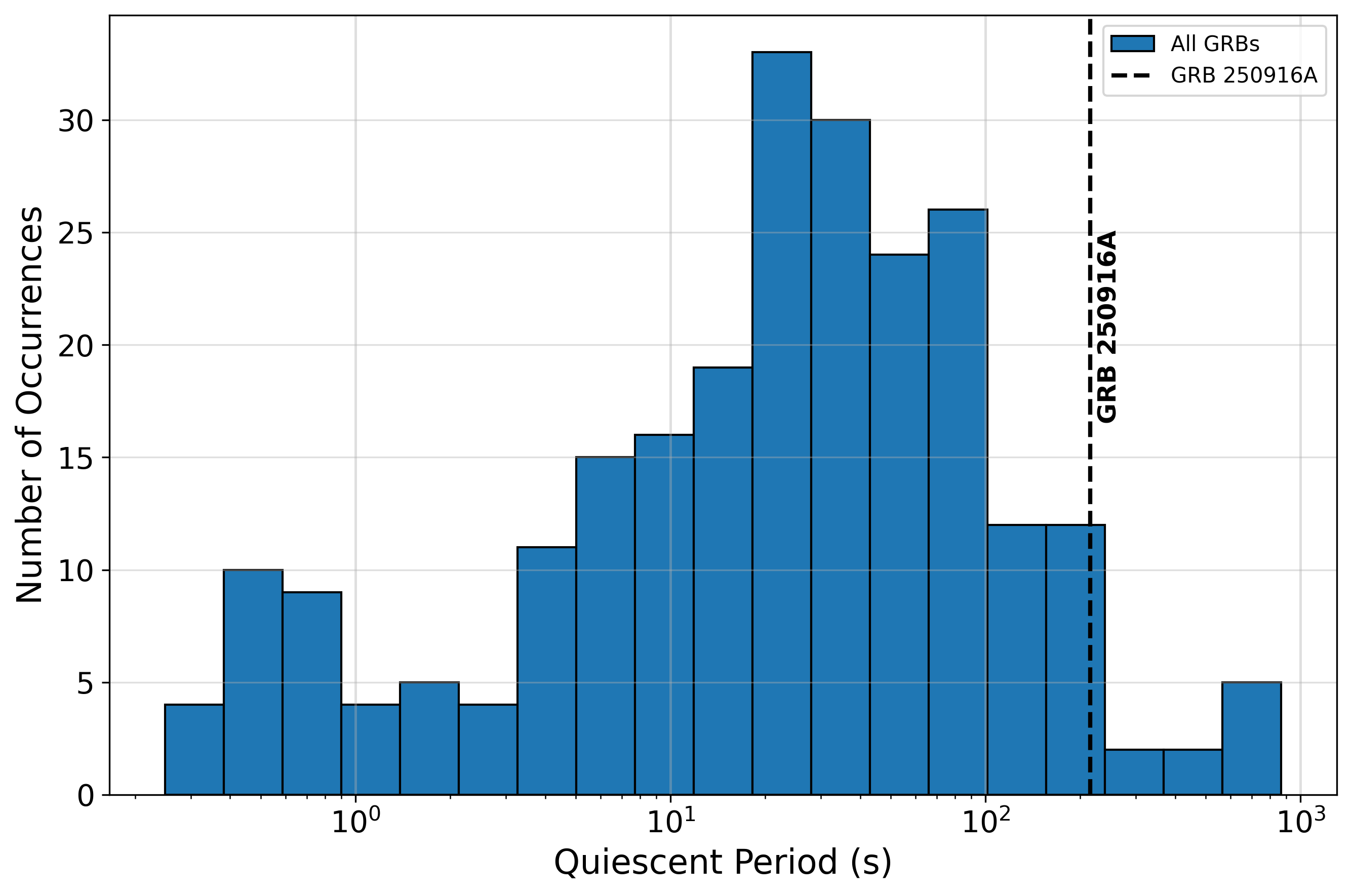}
    \caption{Quiescent time plot for the precursor, plot recreated from~\citet{coppin2020}'s work. \thisgrb\ is at the end of the higher quiescent time tail.}
    \label{fig: quiescent time plot}
\end{figure}

The quiescent interval of $\sim150$~s observed in \thisgrb\ lies at the high end of the long-gap component (Figure~\ref{fig: quiescent time plot}). However, the spectral properties of \thisgrb\ do not align cleanly with either scenario. The precursor is distinctly thermal, consistent with photospheric or shock-breakout emission, whereas the main episode is nonthermal and well described by a Band spectrum. The strong spectral contrast disfavors a simple engine turn-off scenario in which both episodes arise from the same emission mechanism.

As discussed in Section~\ref{sec:cocoon}, the thermal precursor is naturally explained as emission from a cocoon formed during jet propagation through the progenitor star, while the subsequent nonthermal main emission originates from the relativistic jet observed off-axis. This interpretation is spectrally consistent with the data. Nevertheless, purely geometric or breakout-related delays are expected to produce quiescent intervals of up to a few tens of seconds \citep{morsony2007}, shorter than the observed $\sim150$~s.

We therefore infer that the large quiescent interval in \thisgrb\ likely results from a combination of processes: cocoon shock breakout producing the thermal precursor, followed by delayed jet activity associated with central engine intermittency. 

\subsection{Thermal Origins of the Precursor}\label{sec:cocoon}

As shown in Section~\ref{sec:prompt}, the precursor emission in \thisgrb\ lasts for $\sim25$~s and is well described by a thermal blackbody spectrum with a temperature of $kT = 13.2^{+1.5}_{-2.1}$~keV, consistent with the range of temperatures reported for thermal precursors in long GRBs~\citep{ryde2005,burlon2008}. The absence of a fast-rise exponential-decay (FRED) morphology for the precursor disfavors a purely photospheric origin~\citep{Daigne2002,li2007}. The thermal spectral shape and duration are consistent with theoretical expectations for shock breakout from a cocoon formed during the propagation of the jet through the progenitor star~\citep{ramirez2002,nakarpiran2017,colle2017}. Emission from a cocoon is expected to be weakly beamed and quasi-isotropic compared to the main jet~\citep{ramirez2002,colle2017}. The inferred isotropic-equivalent energy of the precursor, $E_{iso} = (9.6 \pm 0.9)\times10^{51}$~erg, lies well within the predicted energy budget of cocoon emission, $E_{\mathrm{iso}} \sim 10^{51}$--$10^{52}$~erg~\citep{nakarpiran2017}.

The presence of a cocoon and the associated pressure exerted during jet breakout can efficiently collimate the relativistic outflow~\citep{lazzati2005,morsony2007}. This mechanism naturally explains the narrow jet half-opening angle of $\theta_c = 0.8^{+0.2}_{-0.1}$~deg inferred from the afterglow modeling. Such a structured jet with a narrow core is consistent with numerical simulations of jet–cocoon systems \citep{gottlieb2021}.    

In conclusion, the thermal precursor, the long quiescent interval, and the narrow structured jet geometry observed in \thisgrb\ can be consistently interpreted within a jet–cocoon framework, coupled with a probable main engine turn-off. The thermal precursor and its energetics favor an origin in cocoon shock breakout, while the extended quiescent phase is consistent with off-axis viewing geometry. The strong collimation inferred from the afterglow modeling further supports the role of cocoon pressure in shaping the jet structure. \thisgrb\ therefore provides a compelling example in which prompt emission, quiescence, and afterglow properties jointly constrain the jet-launching and propagation processes in long GRBs.

\begin{acknowledgments}
We thank all members of the GROWTH collaboration for helping with observations and data processing.

The GROWTH India Telescope \citep[GIT; ][]{2022AJ....164...90K} is a 70-cm telescope with a 0.7-degree field of view, set up by the Indian Institute of Astrophysics (IIA) and the Indian Institute of Technology Bombay (IITB) with funding from DST-SERB and IUSSTF. It is located at the Indian Astronomical Observatory (Hanle), operated by IIA. We acknowledge funding by the IITB alumni batch of 1994, which partially supports the operations of the telescope. Telescope technical details are available at \url{https://sites.google.com/view/growthindia/}.

This work is partially based on data obtained with the 2m Himalayan Chandra Telescope of the Indian Astronomical Observatory (IAO). We thank the staff of IAO, Hanle, and CREST, Hosakote, that made these observations possible. The facilities at IAO and CREST are operated by the Indian Institute of Astrophysics, Bangalore.

G.C. Anupama acknowledges support from the Indian National Science Academy (INSA) under its Senior Scientist programme.

We thank Nikhil Sarin for important inputs on the quiescent interval between the main emission and the precursor emission episodes. 

M.W.C. acknowledges support from the National Science Foundation with grant numbers PHY-2117997, PHY-2308862 and PHY-2409481.

M.B. is supported by a Wübben Stiftung Wissenschaft Student Grant. Funded in part by the Deutsche Forschungsgemeinschaft (DFG, German Research Foundation) under Germany's Excellence Strategy – EXC-2094/2 – 390783311. This paper contains data obtained at the Wendelstein Observatory of the Ludwig-Maximilians University Munich. We thank Christoph Ries carrying out the observations.

Based on observations obtained with the Samuel Oschin Telescope 48-inch and the 60-inch Telescope at the Palomar Observatory as part of the Zwicky Transient Facility project. ZTF is supported by the National Science Foundation under Award \#2407588 and a partnership including Caltech, USA; Caltech/IPAC, USA; University of Maryland, USA; University of California, Berkeley, USA; Cornell University, USA; Drexel University, USA; University of North Carolina at Chapel Hill, USA; Institute of Science and Technology, Austria; National Central University, Taiwan, and the German Center for Astrophysics (DZA), Germany. Operations are conducted by Caltech's Optical Observatory (COO), Caltech/IPAC, and the University of Washington at Seattle, USA.

SED Machine is based upon work supported by the National Science Foundation under Grant No. 1106171. We thank Anna Ho for triggering P60 for observing this GRB.

This work made use of data supplied by the UK Swift Science Data Centre at the University of Leicester.

CZT--Imager is the result of a collaborative effort involving multiple institutes across India. The Tata Institute of Fundamental Research in Mumbai played a central role in spearheading the instrument's design and development. The Vikram Sarabhai Space Centre in Thiruvananthapuram contributed to electronic design, assembly, and testing, while the ISRO Satellite Centre (ISAC) in Bengaluru provided expertise in mechanical design, quality consultation, and project management.
The Inter University Centre for Astronomy and Astrophysics (IUCAA) in Pune was responsible for the Coded Mask design, instrument calibration, and the operation of the Payload Operation Centre. The Space Application Centre (SAC) in Ahmedabad supplied the essential analysis software, and the Physical Research Laboratory (PRL) in Ahmedabad contributed the polarization detection algorithm and conducted ground calibration. Several industries were actively involved in the fabrication process, and the university sector played a crucial role in testing and evaluating the payload.
The Indian Space Research Organisation (ISRO) not only funded the project but also provided essential management and facilitation throughout its development.
\end{acknowledgments}

\begin{contribution}

All authors contributed equally to this collaborative work.

\end{contribution}

%
\facilities{\fermi\ (GBM), \swift\ (XRT), GIT:0.7\,m, HCT:2\,m, FTW:2.1\,m (3KK), ZTF, SEDM, AMI.}

\software{Astropy~\citep{2013A&A...558A..33A,2018AJ....156..123A,2022ApJ...935..167A}, Source Extractor~\citep{1996A&AS..117..393B}, Astro-SCRAPPY~\citep{2019ascl.soft07032M}, \sw{solve-field} astrometry engine~\citep{2010AJ....139.1782L}, \sw{PSFEx}~\citep{2013ascl.soft01001B}, \texttt{SCAMP}~\citep{2006ASPC..351..112B},
\texttt{SWarp}~\citep{2002ASPC..281..228B}, \texttt{photutils}~\citep{Bradley2024}, \texttt{dustmaps}~\citep{2018JOSS....3..695M}, \jetsimpy~\citep{2024ApJS..273...17W}, \sw{PyMultiNest}~\citep{2014A&A...564A.125B}, \sw{3ML}~\citep{2015arXiv150708343V}.}

\appendix
\section{Afterglow Data}
\label{appendix:afterglow}

Here, we describe the data used to analyze \thisgrb, including the reduction and calibration procedures.

\subsection{GIT}
We used the GROWTH-India Telescope (GIT), a 0.7~m fully robotic telescope at the Indian Astronomical Observatory (IAO), Hanle--Ladakh, to obtain early optical observations of the \thisgrb\ afterglow~\citep{2025GCN.41858}. The field was imaged in the Sloan \rband\ filter beginning at $28.08$~hours post-trigger. Data were processed using the automated GIT reduction pipeline, including bias subtraction, flat-fielding, and cosmic-ray removal via \texttt{Astro-SCRAPPY}~\citep{2019ascl.soft07032M}. Astrometric calibration used the offline \texttt{solve-field} engine~\citep{2010AJ....139.1782L}, and photometry was extracted using \texttt{SExtractor}~\citep{1996A&AS..117..393B} and \texttt{PSFEx}~\citep{2013ascl.soft01001B}. Zeropoints were derived relative to the Pan-STARRS DR1 catalog~\citep{2016arXiv161205560C}. The afterglow was detected with AB magnitude \rband$= 19.73 \pm 0.06$ at $28.13$~hours after the trigger.

\subsection{Swift/XRT}
\label{appendix:xrt}
The Neil Gehrels Swift Observatory X-Ray Telescope (\swift/XRT) initiated follow-up observations of \thisgrb\ at $T_0 + 101900$~s ($\approx 1.18$~days) after the \fermi/GBM trigger~\citep{2025GCN.41862}. A total exposure of 1.6~ks was obtained in photon-counting (PC) mode and processed using the automatic analysis pipeline provided by the UK Swift Science Data Centre to obtain the XRT light curves, spectra~\citep{evans2009}. A flux of $(9.2 \pm 1.6)\times10^{-13}~\mathrm{erg\ cm^{-2}\ s^{-1}}$ was measured in the \(0.3-10\)~keV band. A second observation was performed at $T_0 + 781948$~s ($\approx 9$~days), yielding upper limits of $2.0\times10^{-13}~\mathrm{erg\ cm^{-2}\ s^{-1}}$ in the \(0.3-10\)~keV band and $ \leq 2.36\times10^{-6}~\mathrm{erg\ cm^{-2}\ s^{-1}}$ flux density at the 10~keV. 
 
\subsection{FTW}
The Fraunhofer Telescope at Wendelstein Observatory (FTW; \citealt{FTW}) observed the afterglow with the Three-Channel Imager (3KK; (GRBs; \citealt{3KK})) on three consecutive nights in the $r$, $i$, and $J$ bands simultaneously. The first epoch was taken starting at 2025-09-19T23:23:41 UT, approximately 3.4 days post-trigger~\citep{2025GCN.41921}. The optical CCD and NIR CMOS data were reduced with a custom pipeline developed at Wendelstein Observatory, which includes bias, dark, and flat-field corrections, bad-pixel removal, and non-linearity corrections for the CMOS data~\citep{2002A&A...381.1095G, 2025A&A...701A.225B}. The afterglow is detected in all three nights (see Table~\ref{tab:afterglow_obs_table}). The astrometric solution was computed with sources from the Gaia EDR3 catalog~\citep{Gaia2021, 2021A&A...649A...2L, gaiaEDR3} and the photometry was calibrated against the PS1 catalog (optical bands; \citealt{Chambers2016}) and the 2MASS catalog ($J$ band; \citealt{Skrutskie2006}).

\subsection{HCT}
 We observed the field of \thisgrb\ using the Himalayan Faint Object Spectrograph Camera mounted on the 2~m Himalayan Chandra Telescope (HCT) at the Indian Astronomical Observatory (IAO), Hanle, India. Observations were carried out in the \rband\ band, with the middle time of the observation at 2025-09-24T22:50:11.047~UT, $\sim8$~days post-trigger. A total exposure time of 2100~s  was obtained. Standard data reduction and photometric analysis were performed using Astro-SCRAPPY, \sw{SExtractor}, the offline astrometry.net algorithm, and \sw{PSFEx}, as in the case of GIT. The magnitudes are calibrated against PanSTARRS DR1. A $5-\sigma$ upper limit of magnitude \rband$\gtrsim 23.1$ was derived.

\subsection{ZTF}
The optical afterglow was followed by the Zwicky Transient Facility (ZTF; \citealt{2019PASP..131a8002B, 2019PASP..131g8001G, 2020PASP..132c8001D, 2019PASP..131a8003M}) in the SDSS \rband, \gband. ZTF is a public-private survey that images the entire northern sky over a cadence of two days. ZTF deploys the Kowalski alert stream for finding detections that are coincident with the location of the \fermi/GBM trigger. These were filtered for pre-detections, galactic stars, and rocks. The observation tiling was done using \texttt{gewmopt} on this region. The afterglow source was identified and named as \texttt{ZTF25abrvviu}, with the first detection around \(19.25\)~hours.
 
\subsection{SEDM}
Once the afterglow was identified, we acquired images with the Spectral Energy Distribution Machine (SEDM; \citealt{2018PASP..130c5003B,2019A&A...627A.115R}), operating on the Palomar 60-inch telescope. We acquired images in the SDSS \rband\ and \iband, $\sim 3.5~$days after the trigger. The SEDM images were processed with a python-based pipeline version of \texttt{Fpipe} (Fremling Automated Pipeline; \citealt{2016A&A...593A..68F}), which includes photometric calibrations. We measured the afterglow with $r=21.50\pm0.21$, $i=21.17\pm0.19$, before correcting for galactic extinction.

\subsection{AMI}
The Acminute Microkelvin Imager radio telescope team observed the position of \thisgrb\ $\sim 15$~days post trigger. A nearby bright contaminating source, also seen in NVSS, limits the effective sensitivity at the GRB location. No emission is detected at the GRB position, and it is conservatively estimated that any source brighter than $\sim 1$~mJy would have been detected despite the $\sim4$~mJy neighbour. An upper limit of $\lesssim 1$~mJy on the millimetre-wave flux density at 15.5~GHz band is therefore placed on \thisgrb.

\startlongtable
\begin{deluxetable*}{cccccccc}
\tablecaption{Multi-wavelength afterglow observations of \thisgrb\ from optical, X-ray and radio band. The table includes the time since the burst ($T-T_0$) in seconds, filter or band, central frequency (in Hz), measured magnitude (in AB system), unabsorbed flux density (in mJy), galactic extinction corrected magnitude (in AB system)~\citep{2011ApJ...737..103S}, and observing instrument, along with references for each data point.} \label{tab:afterglow_obs_table}
\tabletypesize{\scriptsize}
\tablehead{
\colhead{Time - $T_0$} & \colhead{Filter} & \colhead{Frequency} & \colhead{Mag} & \colhead{Flux} & \colhead{Corr Mag} & \colhead{Instrument} & \colhead{Ref.} \\
\colhead{(sec)} & \colhead{} & \colhead{($\times 10^{14}$ Hz)} & \colhead{AB} & \colhead{mJy} & \colhead{AB} & \colhead{} & \colhead{}
}
\startdata
37548 & L & 5.55516 & 18.90$\pm$0.11 & - & - &  GOTO & \cite{2025GCN.41847} \\
47988 & L & 5.55516 & 19.17$\pm$0.07 & - & - & GOTO & \cite{2025GCN.41847} \\
52092 & L & 5.55516 & 19.28$\pm$0.12 & - & - &  GOTO & \cite{2025GCN.41847} \\
106164 & R & 4.67914 & 19.76$\pm$0.03 & - & 19.65$\pm$0.03 & MAO/AZT-22 & \cite{2025GCN.41860} \\
139104 & r & 4.81208 & 20.00$\pm$0.10 & - & 19.88$\pm$0.10 &  OSIRIS+ & \cite{2025GCN.41863} \\
71208 & R & 4.67914 & 19.56$\pm$0.10 & - & 19.45$\pm$0.10 & KAIT & \cite{2025GCN.41873} \\
156132 & R & 4.67914 & 20.31$\pm$0.10 & - & 20.2$\pm$0.10 & KAIT & \cite{2025GCN.41873} \\
192744 & R & 4.67914 & 20.52$\pm$0.02 & - & 20.41$\pm$0.02 & Mondy & \cite{2025GCN.41892} \\
294849 & r & 4.81208 & 21.75$\pm$0.06 & - & 21.59$\pm$0.06 & FTW-3KK& \cite{2025GCN.41921} \\
294849 & i & 3.92913 & 21.34$\pm$0.06 & - & 21.20$\pm$0.06 & FTW-3KK& This work \\
294849 & J & 2.40161 & 20.62$\pm$0.11 & - & 20.52$\pm$0.11 & FTW-3KK& This work \\
372717 & r & 4.81208 & 21.94$\pm$0.09 & - & 21.87$\pm$0.09 & FTW-3KK& This work \\
372717 & i & 3.92913 & 21.90$\pm$0.12 & - & 21.80$\pm$0.12 & FTW-3KK& This work \\
372272 & J & 2.40161 & 20.87$\pm$0.17 & - & 20.75$\pm$0.17 & FTW-3KK& This work \\
458218 & r & 4.81208 & 23.14$\pm$0.30 & - & 22.97$\pm$0.30 & FTW-3KK& This work \\
458218 & i & 3.92913 & 22.25$\pm$0.20 & - & 22.17$\pm$0.20 & FTW-3KK& This work \\
458230 & J & 2.40161 & $>$ 21.39 & - & - & FTW-3KK& This work \\
101628 & r & 4.81208 & 19.80$\pm$0.06 & - & 19.68$\pm$0.06 & GIT & This work \\
103860 & i & 3.92913 & 19.75$\pm$0.07 & - & 19.66$\pm$0.07 & GIT & This work \\
184932 & r & 4.81208 & 20.69$\pm$0.10 & - & 20.57$\pm$0.10 & GIT & This work \\
102744 & g & 6.28496 & 20.17$\pm$0.06 & - & 20$\pm$0.06 & GIT & This work \\
295884 & R & 4.67914 & 21.45$\pm$0.02 & - & 21.34$\pm$0.02 & CrAO & \cite{2025GCN.41936} \\
387396 & R & 4.67914 & 21.89$\pm$0.03 & - & 21.78$\pm$0.03 & CrAO & \cite{2025GCN.41936} \\
470519 & R & 4.67914 & 22.72$\pm$0.05 & - & 22.61$\pm$0.05 & CrAO & \cite{2025GCN.41936} \\
69282 & g & 6.28496 & 20.02$\pm$0.09 & - & 19.85$\pm$0.09 & ZTF & This work \\
69364 & g & 6.28496 & 20.10$\pm$0.08 & - & 19.93$\pm$0.08 & ZTF & This work \\
72802 & r & 4.81208 & 19.58$\pm$0.13 & - & 19.46$\pm$0.13 & ZTF & This work \\
313110 & r & 4.81208 & 21.50$\pm$0.21 & - & 21.38$\pm$0.21 & SEDM & This work \\
313344 & i & 3.92913 & 21.17$\pm$0.19 & - & 21.08$\pm$0.19 & SEDM & This work \\
104688 & u & 8.65202 & 21.52$\pm$0.20 & -  & 21.3$\pm$0.2 & \swift/UVOT & \cite{2025GCN.41938} \\
102600 & white & 7.73534 & 21.90$\pm$0.30 & - & - & \swift/UVOT & \cite{2025GCN.41938} \\
575424 & VT\_R & 3.63385 & 23.00$\pm$0.25 & - & 22.92$\pm$0.25 & \emph{SVOM}/VT & \cite{2025GCN.42024} \\
575424 & VT\_B & 5.45077 & 23.50$\pm$0.30 & - & 23.36$\pm$0.30 & \emph{SVOM}/VT & \cite{2025GCN.42024} \\
724850 & r & 4.81208 & $>$ 23.1 & - & - & HCT & This work \\
102568 & X-ray & 24180.0 & - & $(1.2 \pm 0.3) \times 10^{-5}$ & - & \swift/XRT & \cite{2025GCN.41862} \\
106866 & X-ray & 24180.0 & - & $(1.4 \pm 0.4) \times 10^{-5}$ & - & \swift/XRT & \cite{2025GCN.41862} \\
781948 & X-ray & 24180.0 & - & $<2.4 \times 10^{-6}$ & - & \swift/XRT & This work \\
1296000 & radio & 0.000155 & - & $< 1$ & - & AMI & This work
\enddata
\end{deluxetable*}

\subsection{Other Ground-based Detections}
Additional photometry from the GCN circulars includes detections by GOTO~\citep{2025GCN.41847}, Pan-STARRS, and multiple observatories (e.g. \citet{2025GCN.41873}; \citet{2025GCN.41892}; \citet{2025GCN.41863}). These points, spanning $g$, $r$, $i$, and $J$ bands between 0.7 and 7~days post-burst, were incorporated in our multi-band afterglow analysis (Table~\ref{tab:afterglow_obs_table}).

\bibliography{sample701}{}
\bibliographystyle{aasjournalv7}



\end{document}